# Optical Rectification and Electro-Optic Sampling in Quartz

Vasileios Balos,[1,2] Martin Wolf,[1] Sergey Kovalev,[3] Mohsen Sajadi[1,4, a)]

[1] Fritz Haber Institute of the Max Planck Society, Berlin, Germany
[2] IMDEA Nanociencia, Ciudad Universitaria de Cantoblanco, Madrid, Spain
[3] Helmholtz Zentrum Dresden Rossendorf, Dresden, Germany
[4] Department of Chemistry, University of Paderborn, Paderborn, Germany

a) sajadi@fhi-berlin.mpg.de

**Abstract**.

We report electro-optic sampling (EOS) response and terahertz (THz) optical rectification (OR) in z-cut α-quartz. Due to its small effective second-order nonlinearity, echo-free waveform of intense THz pulses with a few MV/cm electric-field strength can be measured faithfully with no saturation effect. Both its OR and EOS responses are broad with extension up to ~8 THz. Strikingly, the latter responses are independent of the crystal thickness, a plausible indication of strong surface contribution to the total second-order nonlinear susceptibility of quartz. Our study introduces thin quartz plates as the reliable THz electro-optic medium for echo-free, high field THz detection.

**Introduction**.

Electromagnetic radiation with energies in meV range, known as terahertz (THz) radiation is highly relevant for basic science and for diverse technological applications [1–5]. Weak THz waves are typically used for probing and interrogating the low-energy resonances of matter, including rotations and vibrations of molecules [6–10], intraband transitions [11] and charge transport mechanisms [12–14] in semiconductors, energy gaps in superconductors [15,16], collective intermolecular dynamics of liquids [17,18] and biological systems such as DNA [19], proteins [20] and other bio-relevant systems [21–24].

There is also a growing interest to use intense THz pulses for nonlinear spectroscopy [25–28] and for the control of physical properties of matter [12,29–32]. Thanks to the advances in THz technology, short THz pulses with a few MV/cm electric field strength and more than 1 Tesla magnetic-field are routinely available from table-top sources [31,33]. Although, the race in this field is commonly focused on introducing novel media with high efficiency in converting the optical pump power into THz radiation [34–37], an opposite route has to be taken for coherent detection and characterization of intense THz pulses.

To measure the waveform of THz pulses one typically employs electro-optic sampling (EOS) [38–40], an approach based on the Pockels effect for characterizing the waveform of THz pulses with high temporal resolution. However, strong THz fields easily saturate EOS signals in conventional electro-optic media. To avoid this, one needs to reduce the THz field power by introducing additional optics e.g., polarizers or multiple silicon plates [41]. Alternatively, one can use electro-optic media whose nonlinearities are small enough to sustain THz electric fields in MV/cm regime without saturating the measured signals.

In such quest, we study the EOS response as well as the optical rectification (OR) signal of z-cut $\alpha$-quartz (hereafter simply called quartz). The trigonal lattice structure of quartz lacks centrosymmetry [42,43], thereby one may resolve its EOS and OR responses [44]. Notably, the quartz's second order nonlinearity depicted in its effective second-order susceptibility $d_{eff} \approx 1$ pm/V is about two orders of magnitude smaller than that of ZnTe ($d_{eff} = 68.5$ pm/V), a commonly used nonlinear medium for THz generation and its coherent detection. Quartz also has a large indirect electronic band gap, 5.7 eV, accompanied by a large transparency window, ranging from λ≈3.5 μm to λ≈200 nm and it is transparent at the THz frequency range[45]. In addition, quartz is a hard crystalline material with high optical damage threshold, which makes it an excellent candidate for preparing free standing ultra-thin EOS media.

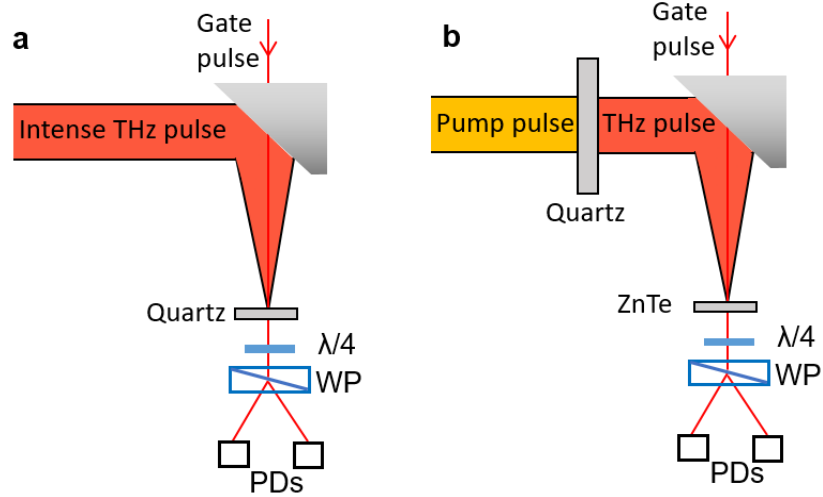

**Fig. 1. Schematic of THz generation and detection setups. a)** Quartz plates with different thicknesses are used as the nonlinear media for electro-optic sampling (EOS) of intense THz pulses. **b)** THz emission from the quartz plates is measured in EOS configuration, where the nonlinear medium is a thin ZnTe crystal. λ/4: quarter wave-plate, WP: Wollaston prism, PDs: two photodiodes for balance detection.

In this contribution we introduce ultra-thin quartz plates as excellent EO media for resolving the waveform of intense THz electric fields, faithfully. Moreover, as quartz is often used as the substrate in THz emission experiments [46,47], we have characterized its THz emission.

**Experiment.**

The details of our experimental setup is given elsewhere. [27,33] Briefly, as schematically shown in **Fig. 1**, quartz plates with different thicknesses are used both as the THz emitter and also electro-optic medium. In **Fig. 1a** quartz plates with different thicknesses are used as EO media. The intense THz pulses are generated either via optical rectification in a Lithium Niobate (LN) crystal, using the pulse tilting approach [41,48], or using a large-scale spintronic THz emitter (STE) [34,49]. The LN THz pulse peaks at ~1THz, with bandwidth of about ~1.5 THz and more than 2 MV/cm electric-field strength. The STE emission on the other hand is very broad, with a bandwidth exceeding 10 THz and electric-field strength of about 300 kV/cm.

**Fig. 1b** depicts the THz emission setup in which quartz plates are used as the active medium. Here, 800 nm short laser pulses (Energy: 4 mJ, duration: 30 fs, 10 mm diameter) are used to pump the quartz plates in a collinear geometry. In this scheme a thin ZnTe crystal (10 µm thick) or GaP crystal (250 µm thick) are used for the EOS of the emitted THz pulses from the quartz plates.

For EOS, temporally delayed probe pulses (Energy: 2 nJ, Wavelength: 800 nm, duration: 8 fs), derived from the seed laser oscillator (Venteon One), are mixed with THz pulses on electro-optic media. The probe pulses are linearly polarized, parallel to the pump polarization, before the sample and subsequently acquire ellipticity owing to the birefringence induced by the co-propagating THz pulse. The ellipticity is detected with a combination of a quarter-wave plate and a Wollaston prism which splits the incoming beam in two perpendicularly polarized beams with power $P_1$ and $P_2$. The normalized difference $(P_1 - P_2)/(P_1 + P_2)$ is twice the probe ellipticity and measured by two photodiodes as a function of temporal delay between THz pump and optical probe pulse. The quartz plates with thicknesses 5 mm, 1 mm, 0.5 mm, 100 $\mu$m, 50 $\mu$m are all z-cut and acquired commercially. The thinner 3 $\mu$m thick plate is prepared by polishing the 50 $\mu$m thick quartz plate.

**Results.**

*THz emission.* Here we consider the THz emission from quartz crystals with different thicknesses. **Fig. 2** encompasses the major results. The top panel of **Fig. 2a** shows the emission from 5 mm thick quartz. Two well separated THz emissions, delayed by ~9.3 ps are resolved. The middle panel shows the THz emission from 1 mm thick quartz, again with two separated pulses with 1.8 ps time delay. However, for the thinnest quartz plate, 3 μm thick, only one THz pulse can be discerned. The Fourier spectra of the THz pulses from 5 mm thick quartz (pulse 1) and the emission from 3 μm thick quartz are given in **Fig. 2b**. Notably the two spectra display the same profile, where the sharp peak at ~4 THz refers to the low-frequency IR-active phonon mode of quartz.[45] The fluence dependence of the observed THz signal from the thin quartz plate is given in the inset in the lowest panel of **Fig. 2a** and shows linear dependence of the emitted THz with 800 nm pump power.

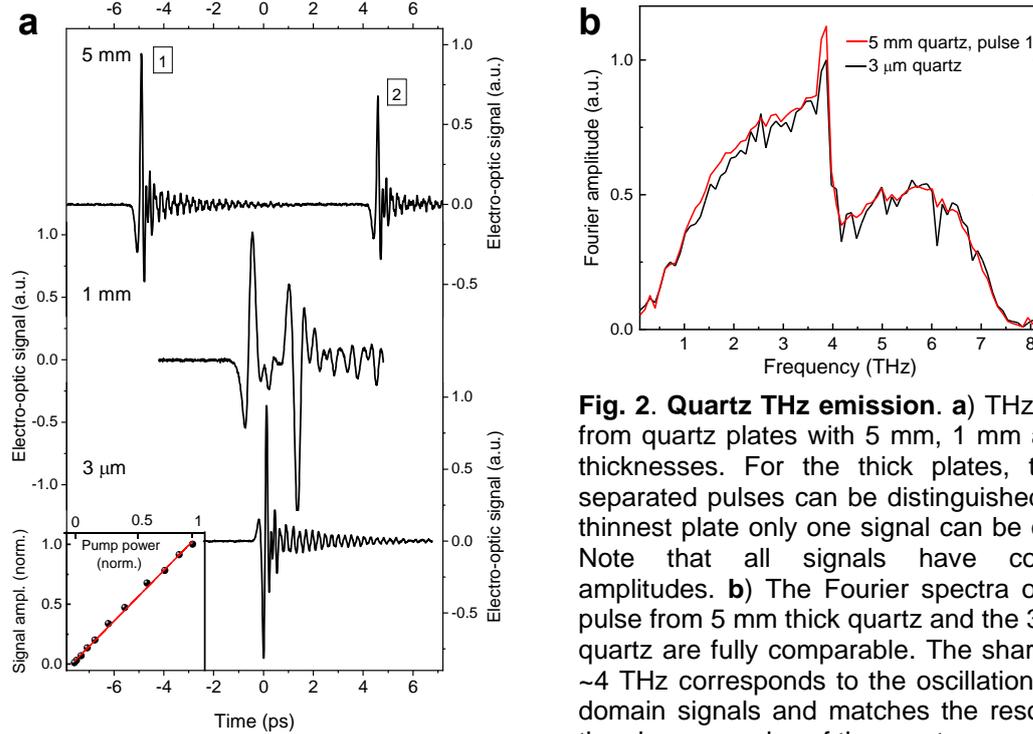

**Fig. 2. Quartz THz emission. a)** THz emission from quartz plates with 5 mm, 1 mm and 3 μm thicknesses. For the thick plates, two time-separated pulses can be distinguished. For the thinnest plate only one signal can be observed. Note that all signals have comparable amplitudes. **b)** The Fourier spectra of the first pulse from 5 mm thick quartz and the 3 μm thick quartz are fully comparable. The sharp peak at ~4 THz corresponds to the oscillations in time-domain signals and matches the resonance of the phonon modes of the quartz.

We have also measured the azimuthal dependence of the THz emission from the quartz plates with thicknesses 5 mm and 3 μm. The polar plots of the resulting signals are shown in **Fig. 3**. All emitted signals show six maxima, the characteristic emission pattern of a system with three-fold symmetry. However, there is a notable phase shift between the radiations from the two radiations observed from the thick plates. In case of the 5 mm and 1 mm thick quartz plates, the azimuthal dependence of the first pulse and the second pulse have close to ~10 degrees and ~3 degrees offset, respectively.

*Electro-optic sampling.* In **Fig. 4a** we show the EOS response of quartz plates with different thicknesses, after pumping the crystals with intense LN THz pulses. For the thickest crystal (5 mm thick) two distinct responses separated by ~9.2 ps can clearly be resolved. For this experiment, the quartz plate is located such that the focus of the THz pulse is close to its center, thereby the two signals appear much broader than the typical LN single cycle THz waveform. For thinner crystals (1 mm and 0.5 mm) also the contribution of two signals can be realized. For thinner crystals <100 μm only a single signal, resembling the waveform of the THz electric field measured with 250 μm thick GaP is resolved, **Fig. 4a** lowest panel, the red line. To avoid saturation of the EOS in GaP, 8 silicon plates, each 200 μm thick are placed in the THz path. Note also that for the thin quartz plates, no reflective echo signals are observed. Due to the large THz wavelength, the reflection effects in ultra-thin plates are all within the main pulse of the THz.

Notably both in the THz emission and in the EOS signals of quartz, the delay of the two signals for different crystal thicknesses cannot be explained by the internal reflection of the THz pulses or the optical pump beam inside quartz. For instance, the echo signals of a THz pulse inside a 5 mm thick quartz should appear at ~66 ps after the first pulse, much longer than the observed ~9.3 ps delay of the observed signals in our experiment.

It is also important to express that there is a puzzling aspect of the EOS and THz emission signals of quartz, namely the amplitude of the latter signals is seemingly and to a good extent independent of the thickness of the quartz plates. However, contrarily, by increasing the effective thickness of the crystals by stacking quartz plates, each with 50 μm thick, we observe that the signal amplitudes are accordingly increased. Here each plate is aligned independently for its maximum EOS response with respect to its azimuthal angle before they are stacked. As displayed in **Fig. 4c**, the amplitude of the signals increases proportional to the number of quartz plates.

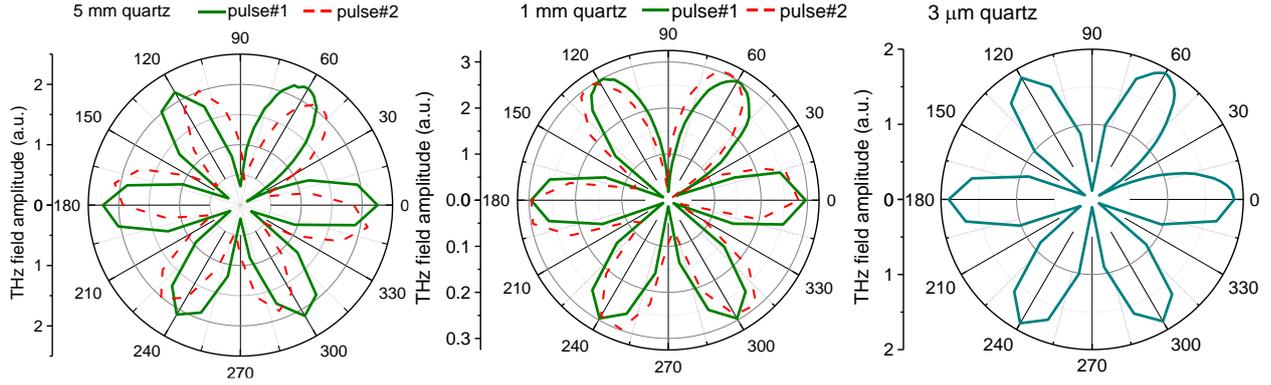

**Fig. 3. Azimuthal dependence in THz emission.** The azimuthal angle (φ) dependence of THz emission from quartz plates with thicknesses 5 mm, 1 mm and 3 μm are shown. While for the thin plate only one signal with six maxima as function of φ is resolved, for thick quartz plates the two delayed signals are phase shifted relative to each other.

We have also measured the azimuthal dependence of the EOS response of quartz, see **Fig. 4b**. Similar to the azimuthal pattern of the quartz plates in THz emission experiment, the resolved signals show six maxima. The solid line is a fit to the experimental data points using $\cos 3\varphi$ as the fitting equation, with $\varphi$ serving as the azimuthal angle.

To obtain the spectral broadness of the EOS response of quartz, we have measured the EOS signal of a thin quartz plate after pumping the crystals with intense THz pulses originating from the STE. **Fig. 5a** and **5b** display the time-trace of the EOS response of quartz and the Fourier spectrum of the measured waveform, respectively. For comparison we have measured the same THz waveform using 10 μm thick ZnTe as the EO medium. Interestingly, the quartz response is extended almost up to 8 THz, with two sharp peaks appearing at 4 THz and 8 THz, emanating from the phonon modes of quartz.[45] The EOS response of thin quartz plates resembles its emission pattern shown in **Fig. 2b**.

*Discussion.* As mentioned above the z-cut quartz has a broken centrosymmetry. The crystal has three-fold rotational symmetry along its c-axis and two-fold symmetry around a-axis. It belongs to the point group 32, Hermann-Mauguin notation [50]. For this symmetry group, crystals have two nonzero tensor elements of the effective second-order susceptibility, which for quartz are $d_{14}= 0.6$ and $d_{11}=1.41$ [51].

The type II phase matching for OR and EOS processes in this symmetry group follows a second-order nonlinearity which depends on the internal phase-matching angle θ and azimuthal angle, this yields $d_{eff} = d_{11} \cos\theta \cos 3\varphi$ [52]. In our experiments, in which the THz wave and the optical beam co-propagate along the c-axis of the quartz plates, the internal angle $\theta = 0$, hence the effective second-order susceptibility is reduced to

$$d_{eff} = d_{11} \cos 3\varphi. \tag{1}$$

This implies that the amplitude of the EOS and OR signals of quartz would appear with six maxima with respect to azimuthal angle $\varphi$ [53–56]. This agrees well with our experimental results presented in **Fig. 3** and **Fig.4b**.

We also employ here a simple analytical model to describe THz emission via OR and EOS signals of quartz. For simplicity we ignore reflections at the crystal surfaces and losses due to the absorption in optical and THz pulses, which are well justified for quartz at optical and THz frequencies. In both OR and EOS processes the phase matching condition in collinear geometry of difference frequency mixing is given by $\Delta K(\Omega, \omega) = K(\omega - \Omega) + K(\Omega) - K(\omega)$, where Ω and ω are, respectively the THz and the optical frequencies. The latter mismatch can be approximated as $\Delta K(\Omega, \omega) \approx \frac{\Omega}{c}\left[n(\Omega) - n_g(\omega)\right]$, where c is the speed of light, $n(\Omega)$ is the refractive index of the medium at THz frequency range, $n_g(\omega) = n(\omega) - \frac{\delta n(\omega)}{\delta \omega}$ is the group refractive index of the nonlinear medium at optical frequencies.

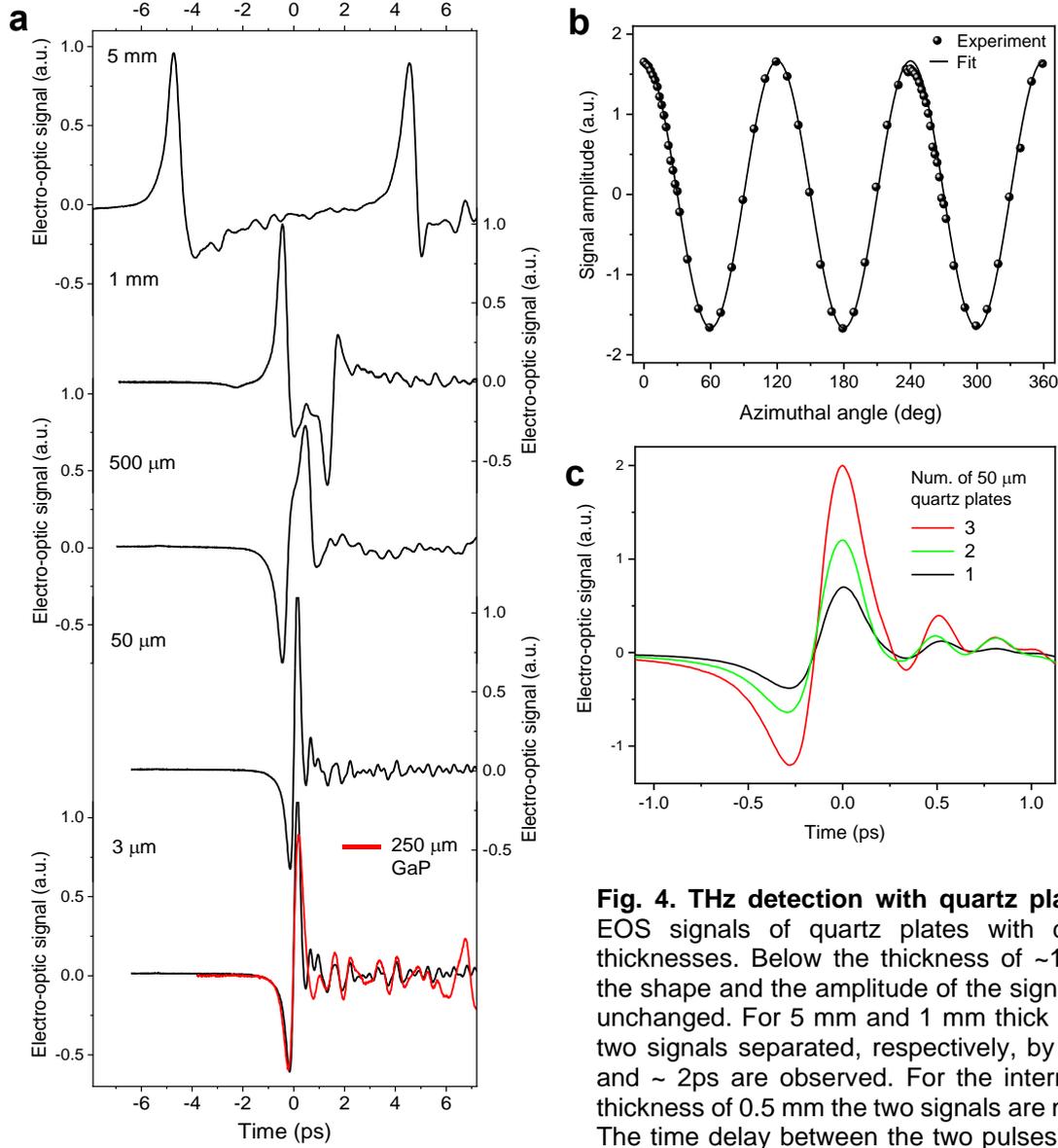

**Fig. 4. THz detection with quartz plates.** **a)** EOS signals of quartz plates with different thicknesses. Below the thickness of ~100 mm the shape and the amplitude of the signals stay unchanged. For 5 mm and 1 mm thick crystals two signals separated, respectively, by ~10 ps and ~ 2ps are observed. For the intermediate thickness of 0.5 mm the two signals are merged. The time delay between the two pulses can be explained by the velocity mismatch between a THz pulse and 800 nm pulse traveling inside quartz plates. **b)** Azimuthal dependence of the EOS signal measured by the 50 μm thick quartz. **c)** While the EOS signal amplitudes of quartz plates in panel **a** are all comparable, the signal amplitude increases proportionally to the number quartz plates (each 50 $\mu$m thick) forming a stack.

As the difference in the refractive indexes at optical THz frequencies of quartz is relatively large, the effective phase matching length, namely its coherence length, $l_c = \pi/\Delta K$ is short and decreases monotonically by increasing THz frequency from $l_c \approx 300~\mu m$ @1 THz to $l_c \approx 50~\mu m$ @5 THz.

For such system, as derived by Gallot and Grischkowsky, the phase factor in OR and EOS processes for a crystal with thickness $d$ is given by [56,57]

$$P(\Omega) \propto i(1 - e^{-i\Delta K(\Omega,\omega)d}), \tag{2}$$

Using Eq. 1, the time profile of the THz pulse can be calculated via the inverse Fourier transform [51,58,59]

$$E_{\text{THz}}(t) \propto \mathcal{F}^{-1}\left[\frac{\Omega^2 \cdot P(\Omega) \cdot d \cdot I(\Omega)}{\Delta K(\Omega,\omega)}\right], \tag{3}$$

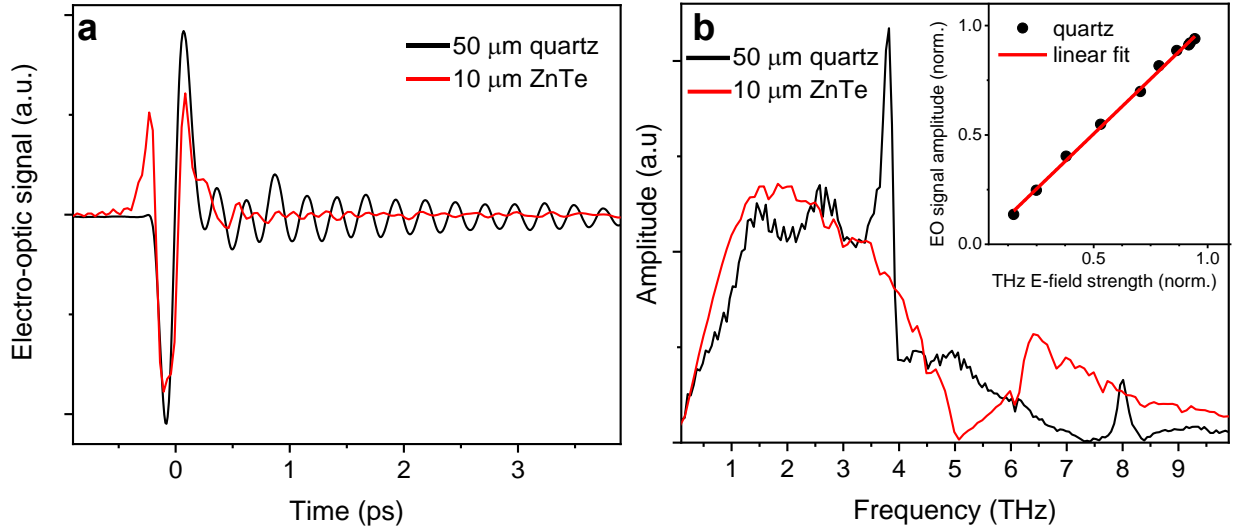

**Fig. 5. Broadband THz detection.** a) Time-domain EOS response of a quartz plate with 70 μm thickness pumped with the emission from a spintronic THz emitter. The THz field strength is about 100KV/cm b) Fourier spectrum of the wave in panel a (black line) is compared to the STE emission measured with 10 μm thick ZnTe crystal (red line). The inset shows the EOS signal measured with quartz scales linearly with the THz pump power.

in which $I(\Omega)= I_0(\Omega)e^{-\frac{\tau^2\Omega^2}{2}}$ is the Fourier spectrum of the optical pulse, assumed to have Gaussian temporal shape with bandwidth $\tau$.

The detected THz electric field via EOS can also be calculated using the same phase factor in Eq. 2. Here the resolved signal can be obtained via [60–62]

$$S(t) \propto \mathcal{F}^{-1}\left[\frac{\Omega^2 P(\Omega) E_{\text{THz}}(\Omega)}{\Delta K(\Omega,\omega)}\right] \qquad (4)$$

Using Eq. 3 and Eq. 4, we have calculated the emitted THz pulse and also the expected signal in the EOS process of quartz with different thicknesses. Interestingly, as shown in **Fig. 6** the time-separated two THz fields observed in thick quartz plates are captured in the calculated signals as well. It is worth mentioning that the two well separated THz emission signals have been observed previously for thick ZnTe, LiNbO$_3$ and LiTaO$_3$ crystals, too [34,63].

Note that the time delay between the observed two signals in thick crystals, matches the time delay between a THz pulse and an optical pulse traversing the width of the crystals. With higher refractive index $n_{\text{THz}} = 2.094$ @1.5 THz, the THz pulse lags behind the optical pulse with refractive index $n_{\text{opt}} = 1.5383$ @800 nm. For instance, the time delay between the two THz pulses generated at the two faces of 5 mm thick crystal will be ~9.2 ps using the relation $\Delta\tau = d\, c^{-1}\Delta n$, where $\Delta n = n_{\text{THz}} - n_{\text{opt}}$ is the difference between refractive indexes of quartz at THz and optical frequencies. As illustrated before by Bakker et al.[Error! Bookmark not defined.] the dual signals in thick crystals can be explained by scrutinizing Eq. 2. The first term in Eq. 2 explains the THz generation and EOS at the front face of crystal and the second term explains the same processes at the rear face of the crystals with the additional phase shift due to the phase mismatch in the generation and detection processes.

Although the main features of the observed experimental results can be explained with this simple model, the observed fairly constant amplitude of the OR and EOS signals remains still puzzling. This is contrary to the expected signals from typical nonlinear media, in which the amplitude of the resolved signals scale with the thickness of the nonlinear medium. A plausible explanation can be the contribution of quartz-air interface to the observed signals. Note that the contribution of bulk to the second-order nonlinear response is typically larger than that of the surface. However, as recently shown in a phase sensitive DFG experiment [64] quartz has a relatively large surface response, comparable with the contribution of its bulk to the DFG response. As such, contribution from the surface may compensate the lower bulk contribution into the resolved signals.

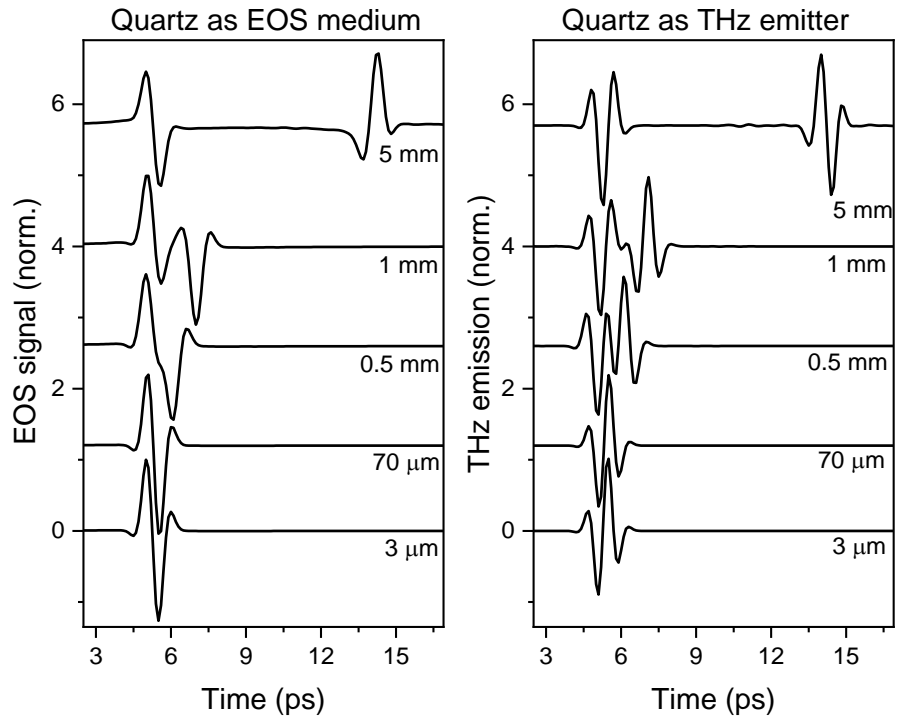

**Fig. 6**.**Optical rectification and Electro-optic sampling in quartz, modeling**. The left panel shows the modeled EOS signals of quartz for different thicknesses. The right panel shows the modeled THz emission from quartz as in left panel. For details see text.

Unfortunately, extracting the surface contribution to the OR and EOS turned out to be challenging. In an effort we modified the surface of the quartz by etching its surface to 10s of nanometer. However, we observed no difference between the EOS signals from etched and non-etched quartz plates, most likely due to failing to turn the crystalline structure of quartz to an amorphous surface by etching.

**Conclusion**.

In this study we have characterized the THz emission and coherent THz detection of intense THz pulses in z-cut α-quartz plates. The emitted THz wave has a broad spectrum extended up to ~8 THz. The same spectral bandwidth is observed in the EOS response of thin α-quartz plates. Importantly, due to the small effective second-order nonlinearity of quartz the echo-free waveform of intense THz electric fields with strength as high as 2 MV/cm can be measured without saturation of the EOS process in α-quartz. Interestingly, due to the large bandgap of quartz, the THz generation and detection can be realized using very wide frequency range of laser pulses from UV to IR.

**Acknowledgements**

Authors would like to thank Mathias Kläui and Gerhard Jakob for valuable discussions, comments on the manuscript and ion etching of a quartz plate. VB acknowledges funding from Marie Skłodowska-Curie Actions programme MSCA-IF-2020-10103087.

**References**

[1] M. Schirmer, M. Fujio, M. Minami, J. Miura, T. Araki, and T. Yasui, Biomed. Opt. Express **1**, 354 (2010).

[2] M.C. Kemp, P.F. Taday, B.E. Cole, J.A. Cluff, A.J. Fitzgerald, and W.R. Tribe, Terahertz Mil. Secur. Appl. **5070**, 44 (2003).

[3] A.Y. Pawar, D.D. Sonawane, K.B. Erande, and D. V. Derle, Drug Invent. Today **5**, 157 (2013).

[4] E. Pickwell and V.P. Wallace, J. Phys. D. Appl. Phys. **39**, R301 (2006).

[5] I. Mehdi, J. V. Siles, C. Lee, and E. Schlecht, Proc. IEEE **105**, 990 (2017).

[6] N. Nagai, R. Kumazawa, and R. Fukasawa, Chem. Phys. Lett. **413**, 495 (2005).

[7] M. Nagai, H. Yada, T. Arikawa, and K. Tanaka, Int. J. Infrared Millimeter Waves **27**, 505 (2006).

[8] C. Rønne, P.-O. Åstrand, and S. Keiding, Phys. Rev. Lett. **82**, 2888 (1999).

[9] M. Takahashi, Crystals **4**, 74 (2014).

[10] S.S. Zhukov, V. Balos, G. Hoffman, S. Alom, M. Belyanchikov, M. Nebioglu, S. Roh, A. Pronin, G.R. Bacanu, P. Abramov, M. Wolf, M. Dressel, M.H. Levitt, R.J. Whitby, B. Gorshunov, and M. Sajadi, Sci. Rep. **10**, 18329 (2020).

[11] B. Sensale-Rodriguez, R. Yan, M.M. Kelly, T. Fang, K. Tahy, W.S. Hwang, D. Jena, L. Liu, and H.G. Xing, Nat. Commun. **3**, 780 (2012).

[12] M. Karakus, S.A. Jensen, F. D'Angelo, D. Turchinovich, M. Bonn, and E. Cánovas, J. Phys. Chem. Lett. **6**, 4991 (2015).

[13] R. Dong, P. Han, H. Arora, M. Ballabio, M. Karakus, Z. Zhang, C. Shekhar, P. Adler, P.S. Petkov, A. Erbe, S.C.B. Mannsfeld, C. Felser, T. Heine, M. Bonn, X. Feng, and E. Cánovas, Nat. Mater. **17**, 1027 (2018).

[14] Z. Mics, K.J. Tielrooij, K. Parvez, S.A. Jensen, I. Ivanov, X. Feng, K. Müllen, M. Bonn, and D. Turchinovich, Nat. Commun. **6**, 7655 (2015).

[15] L. Ozyuzer, A.E. Koshelev, C. Kurter, N. Gopalsami, Q. Li, M. Tachiki, K. Kadowaki, T. Yamamoto, H. Minami, H. Yamaguchi, T. Tachiki, K.E. Gray, W.K. Kwok, and U. Welp, Science. **318**, 1291 (2007).

[16] M. Beck, M. Klammer, S. Lang, P. Leiderer, V. V. Kabanov, G.N. Gol'tsman, and J. Demsar, Phys. Rev. Lett. **107**, 19 (2011).

[17] V. Balos, S. Imoto, R.R. Netz, M. Bonn, D.J. Bonthuis, Y. Nagata, and J. Hunger, Nat. Commun. **11**, 1611 (2020).

[18] G. Schwaab, F. Sebastiani, and M. Havenith, Angew. Chemie - Int. Ed. **58**, 3000 (2019).

[19] A.G. Markelz, A. Roitberg, and E.J. Heilweil, Chem. Phys. Lett. **320**, 42 (2000).

[20] S.J. Kim, B. Born, M. Havenith, and M. Gruebele, Angew. Chemie - Int. Ed. **47**, 6486 (2008).

[21] V. Balos, M. Bonn, and J. Hunger, Phys. Chem. Chem. Phys. **17**, 28539 (2015).

[22] V. Balos, M. Bonn, and J. Hunger, Phys. Chem. Chem. Phys. **18**, 1346 (2016).

[23] V. Balos, H. Kim, M. Bonn, and J. Hunger, Angew. Chemie - Int. Ed. **55**, 8125 (2016).

[24] V. Balos, B. Marekha, C. Malm, M. Wagner, Y. Nagata, M. Bonn, and J. Hunger, Angew. Chemie - Int. Ed. **58**, 332 (2019).

[25] A. Shalit, S. Ahmed, J. Savolainen, and P. Hamm, Nat. Chem. **9**, 273 (2017).

[26] H. Elgabarty, T. Kampfrath, D.J. Bonthuis, V. Balos, N.K. Kaliannan, P. Loche, R.R. Netz, M. Wolf, T.D. Kühne, and M. Sajadi, Sci. Adv. **6**, eaay7074 (2020).

[27] M. Sajadi, M. Wolf, and T. Kampfrath, Nat. Commun. **8**, 14963 (2017).

[28] V. Balos, N.K. Kaliannan, H. Elgabarty, M. Wolf, T.D. Kühne, and M. Sajadi, Nat. Chem. **14**, 1031 (2022).

[29] S.F. Maehrlein, P.P. Joshi, L. Huber, F. Wang, M. Cherasse, Y. Liu, D.M. Juraschek, E. Mosconi, D. Meggiolaro, F. De Angelis, and X.Y. Zhu, Proc. Natl. Acad. Sci. U. S. A. **118**, e2022268118 (2021).

[30] H.A. Hafez, S. Kovalev, J.C. Deinert, Z. Mics, B. Green, N. Awari, M. Chen, S. Germanskiy, U.

Lehnert, J. Teichert, Z. Wang, K.J. Tielrooij, Z. Liu, Z. Chen, A. Narita, K. Müllen, M. Bonn, M. Gensch, and D. Turchinovich, Nature **561**, 507 (2018).

[31] V. Balos, G. Bierhance, M. Wolf, and M. Sajadi, Phys. Rev. Lett. **124**, 093201 (2020).

[32] P. Salén, M. Basini, S. Bonetti, J. Hebling, M. Krasilnikov, A.Y. Nikitin, G. Shamuilov, Z. Tibai, V. Zhaunerchyk, and V. Goryashko, Phys. Rep. **836–837**, 1 (2019).

[33] M. Sajadi, M. Wolf, and T. Kampfrath, Opt. Express **23**, 28985 (2015).

[34] T. Seifert, S. Jaiswal, M. Sajadi, G. Jakob, S. Winnerl, M. Wolf, M. Kläui, and T. Kampfrath, Appl. Phys. Lett. **110**, 252402 (2017).

[35] T. Seifert, S. Jaiswal, U. Martens, J. Hannegan, L. Braun, P. Maldonado, F. Freimuth, A. Kronenberg, J. Henrizi, I. Radu, E. Beaurepaire, Y. Mokrousov, P.M. Oppeneer, M. Jourdan, G. Jakob, D. Turchinovich, L.M. Hayden, M. Wolf, M. Münzenberg, M. Kläui, and T. Kampfrath, Nat. Photonics **10**, 483 (2016).

[36] Y.S. Lee, T. Meade, V. Perlin, H. Winful, T.B. Norris, and A. Galvanauskas, Appl. Phys. Lett. **76**, 2505 (2000).

[37] F. D'Angelo, Z. Mics, M. Bonn, and D. Turchinovich, Opt. Express **22**, 12475 (2014).

[38] Q. Wu and X.C. Zhang, Appl. Phys. Lett. **67**, 3523 (1995).

[39] A. Nahata, D.H. Auston, T.F. Heinz, and C. Wu, Appl. Phys. Lett. **68**, 150 (1996).

[40] A. Nahata, A.S. Weling, and T.F. Heinz, Appl. Phys. Lett. **69**, 2321 (1996).

[41] H. Hirori, A. Doi, F. Blanchard, and K. Tanaka, Appl. Phys. Lett. **98**, 2 (2011).

[42] V.G. Dmitriev, G.G. Gurzadyan, and D.N. Nikogosyan, *Handbook of Nonlinear Optical Crystals* (Springer, Berlin, 1999).

[43] C. Bosshard, K. Sutter, P. Pretre, J. Hulliger, M. Flörsheimer, P. Kaatz, and P. Günter, in *Adv. Nonlinear Opt.*, edited by Gordon and Breach (1995).

[44] F. Zernike and P.R. Berman, Phys. Rev. Lett. **15**, 999 (1965).

[45] C.L. Davies, J.B. Patel, C.Q. Xia, L.M. Herz, and M.B. Johnston, J. Infrared, Millimeter, Terahertz Waves **39**, 1236 (2018).

[46] L. Zhu, Y. Huang, Z. Yao, B. Quan, L. Zhang, J. Li, C. Gu, X. Xu, and Z. Ren, Nanoscale **9**, 10301 (2017).

[47] Z. Fan, M. Xu, Y. Huang, Z. Lei, L. Zheng, Z. Zhang, W. Zhao, Y. Zhou, X. Wang, X. Xu, and Z. Liu, ACS Appl. Mater. Interfaces **12**, 48161 (2020).

[48] J. Hebling, K.-L. Yeh, M.C. Hoffmann, B. Bartal, and K.A. Nelson, J. Opt. Soc. Am. B **25**, B6 (2008).

[49] V. Balos, P. Müller, G. Jakob, M. Kläui, and M. Sajadi, Appl. Phys. Lett. **119**, 091104 (2021).

[50] G.L. Breneman, J. Chem. Educ. **64**, 216 (1987).

[51] T.S. Narasimhamurty, *Photoelastic and Electroßoptic Properties of Crystals* (Plenum Press, New York, 1981).

[52] J.E. Midwinter and J. Warner, Br. J. Appl. Phys. **16**, 1135 (1965).

[53] Y. Huang, Z. Yao, C. He, L. Zhu, L. Zhang, J. Bai, and X. Xu, J. Phys. Condens. Matter **31**, 153001 (2019).

[54] P.C.M. Planken, H.-K. Nienhuys, H.J. Bakker, and T. Wenckebach, J. Opt. Soc. Am. B **18**, 313 (2001).

[55] A. Rice, Y. Jin, X.F. Ma, X.C. Zhang, D. Bliss, J. Larkin, and M. Alexander, Appl. Phys. Lett. **64**, 1324 (1994).

[56] N.C.J. van der Valk, P.C.M. Planken, A.N. Buijserd, and H.J. Bakker, J. Opt. Soc. Am. B **22**, 1714 (2005).

[57] Y.R. Shen, Prog. Quantum Electron. **4**, 207 (1976).

[58] A. Schneider, M. Neis, M. Stillhart, B. Ruiz, R.U.A. Khan, and P. Günter, J. Opt. Soc. Am. B **23**, 1822 (2006).

[59] Z. Wang, IEEE Trans. Geosci. Remote Sens. **1**, 1 (2002).

[60] G. Gallot and D. Grischkowsky, J. Opt. Soc. Am. B **16**, 1204 (1999).

[61] G. Gallot, J. Zhang, R.W. McGowan, T.I. Jeon, and D. Grischkowsky, Appl. Phys. Lett. **74**, 3450 (1999).

[62] A. Leitenstorfer, S. Hunsche, J. Shah, M.C. Nuss, and W.H. Knox, Appl. Phys. Lett. **74**, 1516 (1999).

[63] L. Xu, X.C. Zhang, and D.H. Auston, Appl. Phys. Lett. **61**, 1784 (1992).

[64] M. Thämer, T. Garling, R.K. Campen, and M. Wolf, J. Chem. Phys. **151**, (2019).